\documentclass[a4paper,10pt]{article}
\usepackage{latexsym}
\usepackage{amsfonts,amsmath,amssymb}
\usepackage{graphicx}
\usepackage{color}

\def\eq#1{{Eq.~(\ref{#1})}}
\def\tdot#1{\buildrel\ldots\over{#1}} 

\title{Aspects of electrostatics in a weak gravitational field}
\author{Hamsa Padmanabhan\\
Fergusson College, Pune University, Pune, India\\
\textit{email: hamsa.padmanabhan@gmail.com}\\
 \\
 \\
T. Padmanabhan\\
IUCAA, Pune University Campus, \\
Ganeshkhind, Pune - 411007, India\\
\textit{email: paddy@iucaa.ernet.in}}

\begin{document}

\maketitle

\begin{abstract}
Several features of electrostatics of point charged particles in a weak, homogeneous, gravitational field are discussed using the Rindler metric to model the gravitational field. Some previously known results are obtained by
 simpler and more transparent procedures and are interpreted in an intuitive manner. 
Specifically: (a) We discuss possible definitions of the electric field in curved spacetime (and noninertial frames), argue in favour of a specific definition for the electric field and discuss its properties. (b) We show that the electrostatic potential of a charge at rest in the Rindler frame (which is known and is usually expressed as a complicated function of the coordinates) is expressible as $A_0=q/\lambda$ where $\lambda$ is the affine parameter distance along  the null geodesic from the charge to the field point. (c) This relates well with the result that the electric field lines of a charge coincide with the null geodesics; that is, both light and the electric field lines `bend' in the same manner in a weak gravitational field. We provide a simple proof for this result as well as for the fact that the null geodesics (and field lines) are circles in space. (d) We obtain the sum of the electrostatic forces exerted by one charge on another in the Rindler frame and discuss its interpretation. In particular, we compare the results in the Rindler frame and in the inertial frame and discuss their consistency. (e) We show how a purely electrostatic term in the Rindler frame appears as a radiation term in the inertial frame. (In part, this arises because charges at rest in a weak gravitational field possess additional weight due to their electrostatic energy. This weight is proportional to the acceleration and falls inversely with distance --- which are the usual characteristics of a radiation field.) (f) We also interpret the origin of the radiation reaction term by extending our approach to include a slowly varying acceleration. Many of these results might have possible extensions for the case of electrostatics in an arbitrary static geometry.
\end{abstract}

\section{Introduction and summary}

It is possible to obtain the electromagnetic fields of an \textit{arbitrarily} moving charged particle by first working out the fields in the rest frame of the particle (which will be a noninertial frame) and then transforming to the inertial frame. As is shown in ref.\cite{agtp}, we only need to introduce an instantaneously co-accelerating charge and use the corresponding  Rindler frame to achieve this goal.
While ref.\cite{agtp} succeeded in obtaining this result, the mathematical expressions were rather unwieldy thereby hiding the essential conceptual simplicity.

In this paper we revisit several aspects of the same problem and also investigate in more detail the electrostatics in a weak, homogeneous, gravitational field
represented by a Rindler metric \cite{rindler}. In particular, we study the electric field generated by point charges
\textit{at rest} in this metric and the mutual electro\textit{static} interaction of charges which are at rest in this frame. Since charges which are at rest in the Rindler frame move along uniformly accelerated 
trajectories in the inertial frame, a charge (or a system of charges) at rest in the Rindler frame
 will generate electromagnetic radiation in the inertial frame. (A uniformly accelerated charge \textit{does} emit radiation \cite{radiationacc}, though this issue occasionally comes up for debate  in the literature!) Accordingly,  the forces acting between the charges will 
now arise from both the Coulomb field and the radiation field generated by each charge. On the other hand, in the Rindler frame, the fields and forces are completely static. It is therefore interesting to ask how the \textit{static} forces between the charges in the Rindler frame  transform partially to \textit{radiation fields} in the inertial frame. We will show that, both radiation
as well as the radiation reaction experienced by charged particles have simple interpretations in the Rindler 
frame. En route to this result, we shall also describe several other curious features associated with
the electrostatics of point charges in Rindler spacetime. A summary of our results is given below:

(a) 
It turns out that
there are essentially two
possible definitions of the electric field $\textbf{E}$
in the Rindler frame.  These two definitions differ by a factor $\sqrt{g_{00}}$. We discuss the conceptual
basis for both these definitions in a general curved spacetime (or in curvilinear coordinates; we shall use the phrase `curved spacetime' to generically include curvilinear coordinates in flat spacetime, except when the distinction is important) and present arguments as to why one of them is a better choice
than the other. This discussion is relevant for clarifying some previous contradictions noted in the literature due to the use of different definitions of electric fields.
Given an electromagnetic field tensor $F_{ab}$, which satisfies the covariant Maxwell equations, we will define the electric and magnetic fields as measured by an observer with four-velocity $u^i$ thorough 
two four-vectors $E^i$ and $B^i$ which are given by:
\begin{equation}
\label{defeb1}
E^{i} = u^{j} F^{i}_{\phantom{i}j}; \qquad B^{i} = \frac{1}{2}\epsilon^i_{\phantom{i}jkl} u^j F^{kl}
\end{equation} 
Both $E^i$ and $B^i$ are spatial vectors in the instantaneous rest frame of the 
observer since $u^i E_i = 0 = u^iB_i$. Hence, each has three independent components and the tensor
$F^{ij}$ can be expressed in terms of  them 
as 
\begin{equation}
F^{ij} = u^j E^i - E^j u^i + \epsilon^{ij}\smallskip _{kl}\ u^k B^l
\end{equation}
We will also express Maxwell's equations entirely in terms of the four-vectors $E^i$ and $B^i$ defined above. 

(b)  The electrostatic potential $A_0(\mathbf{r})$ generated by a charge $q$ which is at rest in the 
Rindler frame is well known in the literature (see \cite{rindpot} as well as the papers in \cite{radiationacc}) but has a fairly complicated form. We will show that this 
expression is actually equal to 
\begin{equation}
A_0(\mathbf{r}) = \dfrac{q}{\lambda}\qquad \lambda=\dfrac{1}{g}\tanh (gt_R)
\end{equation}  
where $\lambda$ is the affine distance
along a null geodesic connecting the field point with the source point at the retarded Rindler time $t_R$. (To the extent we know,
this simple interpretation of the potential has not been noticed in the literature before.) 

(c) We also obtain several other related results \cite{gron} in a simpler manner. For example, we show that (i) the electric field lines of a charge are circles which coincide with the (spatial projection of) null geodesics in the Rindler frame and that (ii) the electrostatic potential
can be expressed in a very symmetrical manner (see \eq{nicea0} below). 

(d) We next consider two static charges $q_1$ 
and $q_2$, located at $\mathbf{r}_1$ and $\mathbf{r}_2$, in a weak gravitational field $\mathbf{g}$ and obtain the sum of the mutual electrostatic forces $\mathbf{F}_{12} + \mathbf{F}_{21}$  exerted by  
one charge on the other \cite{sumofforces}. In the absence of the gravitational field, this sum would vanish because the Coulomb forces cancel each other. However, the gravitational field bends the electric field lines downwards in the direction of $\mathbf{g}$ thereby leading to a non-zero value for this sum, given by:  
\begin{equation}
\mathbf{F}_{12} + \mathbf{F}_{21} = \frac{q_1q_2}{lc^2} [\mathbf{g}+(\mathbf{g}\cdot \hat{\mathbf{l}})\hat{\mathbf{l}}] \equiv m_{\rm elec} [\mathbf{g}+(\mathbf{g}\cdot \hat{\mathbf{l}})\hat{\mathbf{l}}]
\end{equation} 
where $\mathbf{l} \equiv \mathbf{r}_1 - \mathbf{r}_2$, 
 the unit vector in the direction of $ \mathbf{l}$ is denoted by $\hat{\mathbf{l}}$ and $m_{\rm elec}=q_1q_2/lc^2$
is the mass corresponding to the electrostatic potential energy.
This result shows that the net force is just $m_{\rm elec} \mathbf{g}$ when
the charges are located in a plane orthogonal to $\mathbf{g}$ but acquires an extra correction
when $\mathbf{g}\cdot \mathbf{l} \ne 0$. We discuss this result 
and explain the physical origin of the extra term by comparing this result with the corresponding
one in the inertial frame. We also show that the result depends crucially on our using the correct
definition of electric field, thereby identifying the source of some previous contradictions \cite{sumofforces, pinto} in the  literature. 

(e) The result described above shows that the electric field of a static charge in the Rindler frame
has a term which is proportional to the acceleration $g$ and inversely proportional to the 
distance from the charge. When transformed to the inertial frame, this field contributes to the standard
radiation term which is proportional to the acceleration and falls inversely with the distance.
It is interesting to note that the weight of the electrostatic energy in the Rindler frame contributes (in part) to
the radiation field in the inertial frame and we  demonstrate and discuss this relation.

(f) By a suitable generalization of the approach described above, it is possible to obtain
the radiation reaction force acting on a charged particle in the non-relativistic limit by
a simple and intuitive method. (This was done earlier in \cite{agtp} for the exact case; by confining ourselves to the weak gravity limit, we can obtain this result in a more transparent manner.)

Given the nature of this topic, we shall try to be  as self-contained as possible and include the pedagogical details in a set of Appendices to the main text. Throughout the discussion, we will use units such that $c = 1$, and we will adopt the signature $(+, -, -, -)$. Latin letters run over 0-3 and Greek letters denote spatial coordinates ranging over 1,2,3.

\section{Rindler frame: Mathematical preliminaries}

The natural coordinate system for discussing a weak, homogeneous, gravitational field is the Rindler coordinate system \cite{rindler} which can be interpreted in term of the coordinate system adopted by a uniformly accelerated observer in flat spacetime.  The metric in the Rindler frame can be expressed in the form:
\begin{equation}
ds^2 = (1 + \mathbf{g}\cdot\mathbf{r})^2 dt^2 - d\mathbf{r}^2
\equiv N^2(\mathbf{r})dt^2 - d\mathbf{r}^2
=(1 + gx)^2 dt^2 - dx^2 - dy^2 - dz^2
\end{equation} 
The second equality defines $N\equiv\sqrt{g_{00}}$ and the form of the metric in the third part is obtained by rotating the spatial coordinates so that the acceleration $\mathbf{g}$ is along the $x-$axis. We note that spatial sections are flat and hence the concept of 3-vector operations in the $t=$ constant surfaces are well-defined by the usual rules of Cartesian vectors. 
 
The transformation equations from the inertial co-ordinates (denoted by capital letters) $(T,\mathbf{R})=(T,X,Y,Z)$, to the Rindler 
co-ordinates $(t,x,y,z)$ are given by $Y = y,Z = z$ and
\begin{equation}
\label{basictr}
gT =  (1 + gx) \sinh (gt); \qquad 1+ gX =  (1 + gx) \cosh(gt)
\end{equation}
which covers the quadrant $|T|<X, (1+gX)>0$ of the inertial frame. Though Rindler-like coordinates can be introduced in other quadrants, this will be adequate for our purpose.
The transformation in \eq{basictr} reduces to an identity (i) when $g=0$, or (ii) at the hypersurface $t=T=0$ even with non-zero $g$. On this hypersurface, $(\partial X^a/\partial x^b)=dia(N,1,1,1)$. These facts will prove to be useful while transforming the tensors from one frame to another. 

We will be often interested in the case of a weak acceleration and work with expressions which are accurate to first order in $g$. In this limit, the  transformations in \eq{basictr} reduce to
\begin{equation}
\label{approxtr}
T\approx t(1+\mathbf{g}\cdot\mathbf{r}); \qquad \mathbf{R}\approx \mathbf{r}+(1/2)\mathbf{g}t^2
\end{equation}
The second relation is obvious; the first one can be interpreted as the effect of gravity on the rate of clocks due to the standard redshift factor. These are correct to linear order in $g$. From \eq{approxtr}, we also have the inverse transformations, again to the lowest order in $g$: 
\begin{equation}\label{approxtr1}
t\approx T(1-\mathbf{g}\cdot\mathbf{R}); \qquad \mathbf{r}\approx \mathbf{R}-(1/2)\mathbf{g}T^2
\end{equation}
Note that to linear order in $g$, we have $\mathbf{g}\cdot\mathbf{R}\simeq\mathbf{g}\cdot\mathbf{r}$. In the Rindler frame, our expressions are correct to $\mathcal{O}(\mathbf{g}\cdot\mathbf{r}/c^2)$ while in the inertial frame, they are correct to order $\mathcal{O}(\mathbf{g}\cdot\mathbf{r}/c^2)$ and $\mathcal{O}(v/c)$, where $v$ is the speed of a particle moving with acceleration $g$.

When we are not interested in the $g\to 0$ limit, it is more convenient to work with a shifted $x-$coordinate $\bar x=x+g^{-1}$ in which the Rindler metric takes the form
\begin{equation}
\label{alt1}
ds^2 
=(g\bar x)^2 dt^2 - d\bar x^2 - dy^2 - dz^2
\end{equation}
with the coordinate transformations:
\begin{equation}
T =  \bar x \sinh (gt); \qquad X =  \bar x \cosh(gt)
\end{equation} 
Curves of constant $\bar x$ correspond to particles travelling on uniformly accelerated trajectories with $t$ being the proper time.

In this form, the transformations reduce to those corresponding to polar coordinates if we analytically continue the time coordinates to purely imaginary values: $t\to it_E; T\to iT_E$. 
This fact is useful while calculating intervals between events and expressing them in terms of Rindler coordinates.
The proper interval between any two events in the Rindler frame can be easily obtained by transforming the corresponding expression in the inertial frame. This can be written down just by inspection if we note that --- when we use the coordinate $\bar x=x+g^{-1}$ and analytically continue to Euclidean space --- the Euclidean distance in the plane between $(t_1^E,x_1)$ and $(t_2^E,x_2)$ is given by the standard cosine formula 
\begin{equation}
s^2_E(2,1)=x_2^2+x_1^2-2x_2x_1\cos g(t_2^E-t_1^E)
\end{equation}
Analytically continuing back and adding the transverse contribution $\rho^2\equiv(y_2-y_1)^2+(z_2-z_1)^2$, we get
\begin{equation}
\label{genssqr1}
s^2(2,1)=\rho^2+ \bar x_2^2+ \bar x_1^2-2 \bar x_2\bar x_1\cosh g(t_2-t_1)
\end{equation}
which will be useful in our discussion later on.   

We will also require the properties  of null geodesics passing through any event in the Rindler frame. If the tangent vector to a null geodesic, emanating from an event $\mathcal{P}$ is $k^a=(\omega, \mathbf{k})$, then we can always choose the transverse coordinates $(y,z)$ such that $\mathbf{k}$ lies in the $xy$ plane. It can be easily verified
\cite{comment1} that null geodesics in this $xy$ plane are parts of circles. (See Appendix \ref{app:nullgeo} for a derivation.) A null geodesic in the $xy$ plane  of the Rindler frame, passing through  $(x=0;y=0)$, say, will be described by the equation
\begin{equation}
(x+g^{-1})^2+(y-y_0)^2=g^{-2}+y_0^2
\label{nullgeo}
\end{equation} 
The affine parameter $\lambda$ for this null geodesic can be obtained from the geodesic equation (see Appendix \ref{app:nullgeo}) and the result is:
\begin{equation}
\lambda=g^{-1}\tanh gt
\label{affine}
\end{equation} 
which is normalized to give $\lambda=t=T$ when $g=0$. We will need these results in what follows.

\section{Definition of electric field in curved spacetime}

Maxwell's equations for the electromagnetic field tensor $F_{ik}=\partial_iA_k-\partial_kA_i$ can be expressed in a generally covariant manner in curved spacetime as:
\begin{equation}
\dfrac{1}{\sqrt{-g}}\partial_{i}(\sqrt{-g}F^{ik}) = 4\pi J^{k}
\label{maxcst}
\end{equation}
where $J^k$ is the current. Once this equation is solved we obtain $F^{ij}$ which, of course, can be transformed covariantly to any other  coordinate system. We would, however, like to introduce a notion of \textit{electric} and \textit{magnetic} fields in a generally covariant manner in the curved spacetime.

In flat spacetime, in inertial coordinates, $F^{\alpha 0},F^\alpha_{\phantom{\alpha}0}$ and $F_{\alpha0}$ differ only in sign and any one of them can be defined as the spatial components of the electric field vector $\mathbf{E}$. This is, of course, not the case in curved spacetime (or even in curvilinear coordinates in flat spacetime) since raising and lowering of the indices will introduce metric components with nontrivial dependence on the coordinates $x^i$. For example, in the Rindler frame, the raising and lowering of the spatial index only changes the sign, but the raising and lowering of the time index introduces the space dependent factor $N^2=g_{00}$. The question arises as to what is the natural and useful definition of the electric field in this context. We shall first provide a general discussion and then specialize to the Rindler frame.

Since even the Lorentz transformation mixes up electric and magnetic fields, it is obvious that we need some extra structure to introduce a generally covariant notion of electric and magnetic fields. One of the simplest choices is to introduce a four-velocity field $u^i$ of an arbitrary observer in the spacetime and define two four-vectors $E^i,B^i$ by the relations:
\begin{equation}
\label{defeb}
E^{i} = u^{j} F^i_{\phantom{i}j}; \qquad B^{i} = \frac{1}{2}\epsilon^{i}_{\phantom{i}jkl} u^j F^{kl}
\end{equation} 
Both $E^i$ and $B^i$ are spatial vectors in the instantaneous rest frame of the 
observer, since $u^i E_i = 0 = u^iB_i$. They together have six degrees of freedom and
$F^{ij}$ can be expressed in terms of  them 
as follows:
\begin{equation}
\label{fijalternate}
F^{ij} = u^j E^i - E^j u^i + \epsilon^{ij}\smallskip _{kl}\ u^k B^l
\end{equation}
It is also easy to verify that, in the instantaneous rest frame of the observer, their spatial components reduce to conventional 
electric and magnetic fields. The dual tensor $*F^{ij}\equiv (1/2)\epsilon^{ijkl}F_{kl}$ is obtained by the replacements $E^i\to B^i,B^i\to -E^i$ in \eq{fijalternate} as one would have expected from the structure of Maxwell's equations.

Our definition of the electric field in \eq{defeb} is motivated by the physical idea that the electric field represents the electromagnetic force per unit charge, experienced by a charge at rest in the spacetime. The contravariant electromagnetic force vector (defined as mass times acceleration) is 
\begin{equation}
mu^j\nabla_ju^a\equiv f^{a} = q F^a_{\phantom{a}b}u^b = q F^{a} \ _{b} \dfrac{dx^b}{ds}
\end{equation}
which leads to the definition of a four-vector $E^i\equiv F^i_{\phantom{i}k}u^k$ as the electric field measured by an observer with four-velocity $u^i$ (which is taken to coincide with the four-velocity of the charge). 

It would be nice to write down  Maxwell's equations directly in terms of $E^i,B^i$ and a unit normalized timelike vector field $u^i$ in the spacetime. (The latter can be thought of as the four-velocities of a congruence of observers in spacetime.).  
By taking the dot products of the Maxwell equations with $u^k$:
 \begin{equation}
u_k\nabla_iF^{ik}=4\pi J^ku_k; \qquad u_k\nabla_i*F^{ik}=0
\end{equation} 
 and manipulating the 
resulting equations (see Appendix \ref{app:maxeb}), we can reduce Maxwell's equations to two scalar equations:
\begin{equation}
\label{max1}
\nabla_i E^i  + a_j E^j +\Omega^l B_l= 4\pi u_j J^j
\end{equation} 
\begin{equation}
\label{max2}
\nabla_i B^i + a_j B^j -\Omega^l E_l=0
\end{equation}
where $a^i\equiv u^j\nabla_j u^i$ is the acceleration of the vector field $u^{i}$, and
$\Omega^l \equiv  \epsilon^{lkij} u_k \nabla_i u_j$ may be thought of as being proportional to the rotation of the observer's velocity field.
We see that the coupling between $E^i$ and $B^i$ is through the term involving $\Omega^l$;
 if  $\Omega_l =0$, then the equations for electric and magnetic fields decouple. 
 Demanding the validity of  \eq{max1} and \eq{max2} \textit{for all four-velocities} $u^i$ leads to the full
set of Maxwell's equations even though \eq{max1} and \eq{max2} are just two scalar equations.

 In static spacetimes, we have a natural choice $u^i=\delta^i_0/N$
for which our definition  leads to an  electric field given by
\begin{equation}
\label{eforce}
E^{i} = F^i_{\phantom{i}k}u^k=\dfrac{F^{i}_{\phantom{i}0}}{N}=NF^{i0}
\end{equation} 
Clearly, only the spatial components of $E^i$ are non-zero and we will take the Cartesian three-vector electric field as the one with the components: $(\mathbf{E})^\alpha=NF^{\alpha 0}(=-N^{-1}F_{\alpha 0})$. We will see, in the sequel, that this definition has several attractive features. 

In this static case  we have  $\sqrt{-g}=N\sqrt{-h}$ where $h$ is the determinant of the spatial metric. The acceleration has only spatial components given by
 $a_\alpha=-\partial_\alpha \ln N$ and $\Omega^l=0$.  Hence, in \eq{max1},
 the $a_iE^i=a_\alpha E^\alpha$  term cancels with a corresponding term involving the derivative of $N$ in $\sqrt{-g}$ in the $\nabla_iE^i$, and the resulting equation simplifies to just $\nabla\cdot \mathbf{E}=4\pi\rho$,
 which is identical in form to that in flat spacetime. 
If the charge distribution is also static, then we can assume that only $A_0$ and $F_{\mu 0}=-F_{0\mu}=\partial_\mu A_0$ are nonzero, leading to $\mathbf{E}=-N^{-1}\nabla A_0$, which is equivalent to the relation $\nabla\times (N\mathbf{E})=0$. So, the
equation we need to solve for the potential in a static geometry is
 \begin{equation}
 \label{maxeqforA}
\nabla\cdot\left(\dfrac{1}{N}\nabla A_0\right)=-4\pi\rho
\end{equation} 
The fact that it is not $\nabla^2A_0=0$ indicates the effect of gravity on the electric field.

For the sake of comparing our results with those in literature, we will also introduce an alternative definition of electric field $\mathbf{E}_{LL}$ whose components are defined simply as $F_{0\alpha}$. (This is the definition used, for example, in Landau-Lifshitz \cite{LL}, which explains the subscript `LL'.) To provide some motivation (which we could not find in the literature) for this definition, consider a static spacetime with a Killing vector
$\xi^{i} = \delta^{i}_{0}$ and $\xi_{i} = N^2 \delta^{i}_{0}$. From the equation of motion for a charged particle
$
mu^a \nabla_a u^{i} = q F^{i} \ _{k} u^{k}
$
we can determine how the energy $mu^i\xi_i$ --- which would have been conserved in the absence of the electromagnetic field --- varies along the trajectory of the particle. We get
\begin{equation}
\dfrac{d\mathcal{E}}{ds}=mu^{a}\nabla_a (u^{i} \xi_{i}) =  q F^{i} \ _{k} \xi_{i} u^{k}
\end{equation}
Expressing $ds$ in terms of $dt$,  this equation can be recast as  
\begin{equation}
\label{ework}
\dfrac{d\mathcal{E}}{dt} = q N (\mathbf{E} \cdot \mathbf{v})
\end{equation}
where $\mathbf{v}=d\mathbf{x}/dt$ is the usual three-velocity.
If we want to interpret the rate of change of energy as due to the work done by the electric field, we can use an alternative definition of the electric field, which we shall call $E^{\alpha}_{LL}$, as
\begin{equation}\label{ene}
E^{\alpha}_{LL} = N E^{\alpha}=N^2F^{\alpha0}=-F_{\alpha0}
\end{equation}
so that the right hand side of \eq{ework} becomes just $q(\mathbf{E}_{LL} \cdot \mathbf{v})$.
In the case of the Rindler frame, the two definitions of electric fields are related by:
\begin{equation}\label{enerind}
E^{\alpha}_{LL} = (1 + \mathbf{g} \cdot \mathbf{r}) E^{\alpha}
\end{equation} 
In static spacetimes, only the $J^0$ and $A_0$ components are relevant and $N\mathbf{E}=\mathbf{E}_{LL}$ will be a spatial gradient ensuring that the work done by the electric force is path independent. Hence $\nabla\times \mathbf{E}_{LL}=
\nabla\times N\mathbf{E}=0$.
Since $\mathbf{E}_{LL}=-\nabla A_0$, the final Maxwell equation for $A_0$ will, of course, be the same as \eq{maxeqforA}, but the equation in terms of the electric field will now be $\nabla\cdot (N^{-1}\mathbf{E}_{LL})=4\pi\rho$, which is sometimes interpreted as curved spacetime acting as an optically active medium with a refractive index.

\textit{The fact that the two definitions of electric field differ by a factor which depends on space coordinates has important implications for electrostatic interaction of charges.} For example, consider the sum of forces exerted by two charges on each other. The result will now depend on the definition we use and --- given the fact that $\mathbf{E}$ is defined through the force equation --- it will be incorrect to use $\mathbf{E}_{LL}$ in this context. \textit{That is, the force acting on a charge is $q\mathbf{E}$ and not $q\mathbf{E}_{LL}$.} This incorrect use is the root cause of some of the contradictory results obtained in the literature \cite{sumofforces, pinto} previously.

The difference between the two definitions also has implications for the  question of  translational invariance of the electric field in a curvilinear coordinate system which does not seem to have been emphasized in previous literature. Suppose we find that the electric field at $\mathbf{r}$ due to a charge at $\mathbf{r}_0$ depends only on $\mathbf{r}-\mathbf{r}_0$ when one definition for the electric field is used. It is then clear that --- in general --- the second definition will not have this property! We will illustrate these features in explicit examples in what follows. 
In most cases, we will use $\mathbf{E}$ based on our definition of the electric field in terms of the force (which we believe leads to physically reasonable results), but we will also quote the corresponding results obtained by using $\mathbf{E}_{LL}$  when it is relevant.

\section{Electric field of a point charge in the Rindler frame}

After these preliminaries we take up the question of determining the potential and the electromagnetic 
field of a charged particle which is at rest in a given point in the Rindler frame or --- equivalently --- in a
weak homogeneous gravitational field. 
Such a charged particle will be moving along a uniformly accelerated trajectory in the inertial
coordinate system. 
The $A_i$ and $F_{ij}$ in the Rindler frame can be obtained  either by  straightforward integration of 
Maxwell's equations or by transforming the corresponding fields in the inertial frame. The resulting 
expressions, well known in literature \cite{rindpot}, appear quite complicated when expressed in Rindler coordinates.
We shall however show  that this result can be obtained in a fairly simple and intuitive manner
and that the final result for the electrostatic potential has an elegant physical interpretation.

We shall first consider a charge at rest \textit{at the origin} of the Rindler frame and obtain its electromagnetic field. We begin by noting that, because of  the static nature of the Rindler frame, the 
vector potential   reduces to the form
 $A_i = (A_0, 0,0,0)$ with $A_0 (\mathbf{r})$ being independent of the time coordinate.
 It is therefore enough if we determine the electrostatic potential on the $t=0$ hypersurface.
We also know that the potential at an event $x^i$ is determined by the nature of the trajectory
 of the charged particle $z^i(t_R)$ at the retarded time $t_R$. This retarded time is a function 
 of the field coordinates $x^i$ and is determined by the condition that $z^i(t_R)$ and 
 $x^i$ are connected by a null geodesic.
We will argue that the potential $A_0 (0, \mathbf{r})$
  due to a charge at rest in the Rindler frame should be expressible in the form
    \begin{equation}
\label{insight}
    A_0 (\mathbf{r})=
A_0 (0, \mathbf{r}) = \frac{q}{\lambda (\mathcal{F}; \mathcal{S})}
\end{equation}
where $\lambda(\mathcal{F}; \mathcal{S})$ is the affine parameter distance
along a null geodesic connecting the field event $\mathcal{F}(0,\mathbf{r})$
with the location of the source at the retarded time $\mathcal{S}(t_R,\mathbf{0})$.

This result is easily established along the following lines: We begin with the usual  
Lienard-Wiechert formula for the potential of an arbitrarily moving charge in inertial coordinates, written in the form:
\begin{equation}
\label{LW}
A_k=\frac{2qu_k}{|ds^2/d\tau|}
\end{equation}
where $u^i(\tau)$ is the four-velocity of the charge in the inertial frame at the proper time $\tau$ and the expression on the right hand side has to be evaluated at the retarded time on the trajectory of the charge (see, e.g., \cite{TPvol1} or Appendix C of \cite{agtp}). Taking the dot product of both sides with $u^k$ (at the retarded time) we get the scalar equation $A_ku^k=2q/|ds^2/d\tau|$. In the Lorentz frame in which the charge was at rest at the origin, at the retarded time,  the right hand side reduces to usual Coulomb form $q/|T_R|$ where $T_R$ ($<0$) is the relevant retarded time satisfying 
 the condition $T_R=-|\mathbf{R}|$. We next note that $-T_R$ or $|\mathbf{R}|$ is 
 actually the affine distance $\lambda$ along the null geodesic connecting the event $\mathcal{S}=(T_R,\mathbf{0})$ corresponding to the source at retarded time to the 
 event $\mathcal{F}= (0,\mathbf{R})$ where the field is measured. 
This shows that we can equivalently write $A_ku^k=q/\lambda$ in any Lorentz frame, for an arbitrarily moving charged particle. But both sides of this equation are also generally covariant in flat spacetime when curvilinear coordinates are used.\footnote{In the left hand side, $A_k$ is the potential at some event $x^i$ while $u^k$ is the four-velocity of the charge at the retarded event $z^i$ connected to $x^i$ by a null geodesic. So the dot product of these two vectors, defined \textit{a priori} in two different events, can be taken only after parallel transporting one vector to the location of another. Since this parallel transport is unique in flat spacetime, the expression is invariant with respect to curvilinear coordinate transformations in flat spacetime. Unfortunately, this prevents us from applying this idea to genuinely curved spacetime without modification.} Therefore we can use the same relation in curvilinear coordinates as well, and express the electrostatic potential of a static 
 source at the origin of the Rindler frame in a generally covariant manner, in terms of the affine parameter
 distance between the source at the retarded time and the field point. In the Rindler frame, we have $A_ku^k=A_0$ since $u^0=1/N=1$ at the trajectory of the charge at all times, including the relevant retarded time, thereby leading to
the result in \eq{insight}.
Since the affine parameter is given by  \eq{affine}, we get the result:
\begin{equation}\label{a0tr}
A_0 (\mathbf{r}) = \frac{q}{g^{-1} \tanh gt_R}
\end{equation}
Obviously, both $\lambda$ and the retarded time $t_R$ depend on 
the spatial coordinate of the field point $\mathbf{r}$. 
The retarded time $t_R$ can be computed quite easily (see Appendix \ref{app:rettime})
and the resulting final expression for the 
electrostatic potential reduces to
 \begin{equation}
A_0 = \dfrac{q}{r}\dfrac{1 + gx + g^2r^2/2}{(1 + gx + g^2r^2/4)^{1/2}}
\label{a0sol}
\end{equation}
where $r^2=x^2+y^2+z^2\equiv\rho^2+x^2$. While this expression has been obtained by several people in the past, the interpretation in terms of the affine parameter --- as far as we know --- was not noticed before.

This result can be expressed (see Appendix \ref{app:rettime}) in a nicer form \cite{gron} in terms of the coordinate $\bar x=x+g^{-1}$ as
\begin{equation}
A_0=\dfrac{qg}{2}\left( \dfrac{\ell_+}{\ell_-}+  \dfrac{\ell_-}{\ell_+}   \right);
\quad\ell_\pm^2=\rho^2+(\bar x\pm g^{-1})^2
\label{nicea0}
\end{equation} 
where $\ell_\pm$ represent the distances to the field point from a charge (at $1/g$) and an `image charge'
(at $-1/g$). Equipotential surfaces correspond to constant values of ${\ell_+}/{\ell_-}$. Since the locus of a point that moves keeping the ratio of distances from two different points constant, is a circle, we find that equipotential surfaces are circles in the $xy$ plane. The Maxwell equation (\eq{maxeqforA}) reduces in our case to
 \begin{equation}
\nabla \cdot \left(\dfrac{\nabla A_0}{(1 + \mathbf{g}\cdot \mathbf{r})}\right) = -4 \pi \rho
=-4 \pi q \delta(\mathbf{r})
\label{a0eqrind}
\end{equation} 
in the Rindler frame with the electric field given by:
\begin{equation}
\label{efroma}
\mathbf{E} = -\dfrac{\nabla A_0}{(1 + \mathbf{g}\cdot \mathbf{r})}
\end{equation} 
One can verify by explicit --- though tedious --- computation that \eq{a0sol} is a solution to \eq{a0eqrind}.

Of course, \eq{a0sol} can also be obtained by the coordinate transformation from the field of a uniformly accelerated charge in the inertial frame. Because this is usually done in a rather complicated manner in the literature (and for the sake of those skeptical of the argument leading to \eq{a0tr}!), we
will provide a proof of \eq{a0tr} by transforming the potential directly from the inertial frame. We begin by noticing that, since the potential in Rindler frame is time independent, it can be conveniently evaluated at $t=T=0$. On this hypersurface, the transformation matrix $\partial X^a/\partial x^b$ has only one nontrivial term $\partial T/\partial t=g\bar x$ and hence the transformation gives $A_0(t=0,\mathbf{r})_{Rind}=g\bar x A_0(T=0,\mathbf{R})_{iner}$.
From the Lienard-Wiechert potential in \eq{LW} we have:
$
A_0^{iner}=2qu_0/|ds^2/d\tau|
$. 
But a charge at
rest in the origin of the Rindler frame has $u_0=\cosh gt_R$ with the Rindler time acting as the proper time. From \eq{genssqr1}, with  $\bar x_1=g^{-1}, y_1=z_1=0, t_1=t, t_2=0, \mathbf{\bar r_2}=\mathbf{\bar r}$, we find $s^2(t) = \bar r^2 + g^{-2} - (2\bar{x}/g) \cosh(gt)$, giving  $|ds^2/dt|=(2\bar x)\sinh gt_R$, at the retarded time $t_R$. Thus we get
\begin{equation}
A_0^{Rind}=(g\bar x) A_0^{iner}=2q \dfrac{g\bar x u_0}{|ds^2/d\tau|}=\dfrac{q}{g^{-1}\tanh gt_R}
\end{equation}
which matches with \eq{a0tr}.
The simplicity of the argument is noteworthy.

The corresponding electric field can be obtained using \eq{efroma}.
Without loss of generality, we can confine our attention to the $xy$ plane with $\mathbf{E}=(E^x, E^y, 0)$. Explicit calculation gives:
\begin{eqnarray}
\label{esol}
E_x &=& \dfrac{qx}{r^3}\dfrac{1 + gx/2 - g y^2/2x}{(1 + gx + g^2r^2/4)^{3/2}}; \nonumber \\ 
E_y &=& \dfrac{qy}{r^3}\dfrac{1 + gx}{(1 + gx + g^2r^2/4)^{3/2}}
\end{eqnarray}
(Since only $A_0$ is nonzero in the Rindler frame, it follows trivially that the magnetic field vanishes identically.) 

One can obtain an intriguing result related to the electric field directly from this expression. We know that the electric field lines in the $xy$ plane are given by curves $x=x(y)$ which satisfy the equation $dx/dy=E^x/E^y$.
On using \eq{esol}, this reduces to
\begin{equation}
\dfrac{dx}{dy}=\dfrac{(x+g^{-1})^2-g^{-2}-y^2}{2y(x+g^{-1})}
\label{efieldline}
\end{equation} 
It is easy to verify that this equation is solved by the circles in \eq{nullgeo} by noting that, for these circles,
\eq{efieldline} gives $dx/dy=-(y-y_c)/(x+g^{-1})$ which is the same relation we get from \eq{nullgeo}.
In other words, \textit{the electric field lines of a static charge in the Rindler frame coincide with the null geodesics!} It is understandable that the electric field lines bend under the action of gravity but it is rather surprising that they do so  exactly like the light rays.
It is not clear (a) whether there is simple way of guessing this result in the case of  Rindler frame and (b) what is the general condition on a spacetime for such a result to hold. (These issues are under investigation e.g., as regards Schwarzschild and De Sitter spacetimes.)

Having obtained the exact results, we shall next consider the case of a weak gravitational field and work out the expressions to the linear order in $g$. (A Rindler frame with acceleration $\mathbf{g}$  corresponds to a weak gravitational field $-\mathbf{g}$ in the  direction opposite to the acceleration; but for simplicity, we shall continue to quote the results in terms of $\mathbf{g}$.)
In this case, \eq{a0eqrind} reduces to
\begin{equation}\label{maxwelllowestorder}
\nabla^2 A_0 \approx - 4 \pi \rho + \mathbf{g} \cdot \nabla A_0  
\end{equation} 
which is correct to linear order in $g$. This equation is easy to solve and we get the solution:
\begin{equation}\label{potloword}
A_0 = \dfrac{q}{r}\left(1 + \dfrac{\mathbf{g}\cdot \mathbf{r}}{2}\right)
\end{equation} 
This is the electrostatic potential, in the  limit of a weak gravitational field, of a charge at rest at the origin of co-ordinates in the Rindler frame. We can use \eq{efroma} to obtain the corresponding electric field from this potential. We get:
\begin{equation}\label{origine}
\mathbf{E} = \dfrac{q \hat{\mathbf{r}}}{r^2} - \dfrac{q}{2r} (\mathbf{g} + (\mathbf g \cdot \mathbf{\hat r})
 \mathbf{\hat r}) 
 =\dfrac{q \hat{\mathbf{r}}}{r^2}\left(1-\dfrac{(\mathbf g \cdot \mathbf{ r})}{2}\right) + \dfrac{q}{2r} (\mathbf{ -g})
\end{equation} where $\mathbf{\hat r}$ denotes the unit vector in the radial direction. (Both these expressions can also be obtained from \eq{a0sol} and \eq{esol} by a Taylor series expansion in $g$.). In the first expression for $\mathbf{E}$ in \eq{origine}, we have given the result in terms of a Coulomb term plus a correction due to the gravitational field. In the second expression, we have separated the two terms based on the direction of the vectors: the first one is in the radial direction with a corrected Coulomb term while the second one is in the direction of the gravitational field ($-\mathbf{g}$).

The form of the potential in \eq{potloword} again has a simple interpretation in terms of the 
retarded time $t_R$. From the exact  expression for the potential in \eq{a0tr}, we see that, to 
linear order in $g$, the potential is given by $A_0 = q/t_R$ (which is exactly the form of the 
potential in inertial coordinates). To the same order of accuracy, $t_R$ can be computed by transforming the condition
$T_R^2=R^2$ in the inertial frame using  \eq{approxtr}. This gives, to linear order in $g$, the result:
\begin{equation}
t_R^2(1+2\mathbf{g}\cdot\mathbf{r})=r^2+t_R^2(\mathbf{g}\cdot\mathbf{r})
\end{equation}
which leads to 
\begin{equation}
t_R \simeq r[1 - (\mathbf{g} \cdot \mathbf{r}/2) ]
\end{equation} 
thereby allowing $A_0 = q/t_R$ to be expressed in the form  in \eq{potloword}. Once again, we find that the weak field expressions have simple physical interpretations.
  
These results are for a charge located at the origin of the Rindler frame. For our applications, we will require the potential and field produced by a charge   at rest, not at the origin, but at an arbitrary point $\mathbf{r}_0=(x_0, y_0,0)$. (As noted before, there is no loss of generality in confining to the $xy$ plane.) 
We cannot simply introduce a translation of coordinates because our background metric is not translationally invariant. To get the correct fields, we will proceed as follows:
We note that, when we make a translation of co-ordinates from $\mathbf{r}$ to $\mathbf{\bar{r}}=\mathbf{r -r_0}$, the equation for determining the potential to the lowest order, \eq{maxwelllowestorder}, remains invariant, but the metric 
changes to
\begin{eqnarray}
ds^2 &=& (1 + \mathbf{g}\cdot \mathbf{r})^2 dt^2 - dx^2 - dy^2 - dz^2  \nonumber \\
 &=& (1 + \mathbf{\bar g}\cdot \mathbf{\bar r})^2 d\bar t^2 - dx^2 - dy^2 - dz^2     
\end{eqnarray} 
 where the new gravitational field is defined by 
 \begin{equation}
\mathbf{\bar g} = \dfrac{\mathbf{g}}{1 +\mathbf{g}\cdot \mathbf{r_0}}\approx \mathbf{g}
\end{equation} 
where the last result is true to the lowest order.
The new time co-ordinate is defined by
\begin{equation}
d\bar t = dt (1 + \mathbf{g} \cdot \mathbf{r_0}).
\end{equation} 
Since the metric in the $(\bar t,\bar{\mathbf{r}})$ coordinate system is in the Rindler form, we already know the potential in these coordinates for a charge located at $\bar{\mathbf{r}}=0$ and this is given by \eq{a0sol} with $\mathbf{r}$ replaced by $\bar{\mathbf{r}}$. That is,

\begin{equation}
\bar A_0=
\dfrac{q}{|\mathbf{\bar r}|}  \left( 1 + \dfrac{\mathbf{g} \cdot \mathbf{\bar r}}{2}\right)=
\dfrac{q}{|\mathbf{r} - \mathbf{r}_0|}  \left( 1 + \dfrac{\mathbf{g} \cdot (\mathbf{r} - \mathbf{r_0})}{2}\right)
\end{equation} 
But a charge at the origin of the $(\bar t,\bar {\mathbf{r}})$ coordinate system corresponds to a charge at $\mathbf{r_0}$ in the original coordinates. To find the potential in the original coordinates, we only have to make the correct transformation obtaining:
\begin{eqnarray}\label{transa}
A_0 &=& \dfrac{\partial \bar t}{\partial  t} \bar A_0 \nonumber \\ 
&\simeq & \left( 1 + \mathbf{g}\cdot \mathbf{r_0}\right)\dfrac{q}{|\mathbf{r} - \mathbf{r}_0|}  \left( 1 + \dfrac{\mathbf{g} \cdot (\mathbf{r} - \mathbf{r_0})}{2}\right) \nonumber \\
  &\simeq & \dfrac{q}{|\mathbf{r} -\mathbf{r}_0|}\left( 1 + \dfrac{\mathbf{g} \cdot (\mathbf{r} + \mathbf{r_0})}{2}\right)
\end{eqnarray} 
where the equalities are correct to the linear order in $\mathbf{g}$.
We see that the potential is \textit{not} translationally invariant with respect to the position of the charge, because it is not a function of the difference $(\mathbf{r} - \mathbf{r_0})$ alone. 
This is not surprising a priori because the background metric breaks the translational symmetry. 

Incidentally, this approach works even in the case of the exact solution (as shown in Appendix \ref{app:trantrick}) and one can determine the potential due to a charge located at $\mathbf{r}_0=(x_0,y_0,0)$ to be
\begin{equation}
A_0 = \dfrac{q}{\mathbf{|r - r_0|}}\dfrac{1 + g(x + x_0) + g^2(x^2 + x_0^2 + (y - y_0)^2)/2}{( 1 + g(x + x_0) + g^2(\mathbf{|r - r_0|})^4/4)^{1/2}}
\end{equation}
Obviously, the exact solution also breaks translational symmetry along the $x-$axis.

What \textit{is} surprising, however, is that  the electric field obtained from the potential in \eq{transa}, by the definition in \eq{eforce} in terms of the force on a charged particle, \textit{does} turn out to be translationally invariant to linear order in $g$. To see this we only have to work out the expressions for the field from \eq{transa} by using the formula \eq{efroma}.
Doing this, we find:
\begin{equation}\label{transe}
\mathbf{E} = \dfrac{q \mathbf{l}}{l^3} - \dfrac{q}{2l} (\mathbf{g} + (\mathbf g \cdot \mathbf{\hat l}) \mathbf{\hat l}); \qquad  \mathbf{l} = \mathbf{r - r_0},
\end{equation} 
 so that this electric field \textit{is} translationally invariant and depends only on the vectorial separation between the charge and the field point.

The fact that our electric field is defined as $\mathbf{E} = -N^{-1}\nabla A_0$ is crucial for this result.
In contrast, the electric field $\mathbf{E}_{LL}$ defined by \eq{enerind}, without the $N^{-1}$ factor by
 simply differentiating the scalar potential with respect to the spatial co-ordinates, is not translationally invariant because $A_0$ is not translationally invariant. Instead, we get:
\begin{equation}\label{ell}
\mathbf{E}_{LL} = \dfrac{q \mathbf{l}}{l^3}\left(1 + \dfrac{\mathbf{g} \cdot (\mathbf{r} + \mathbf{r_0})}{2}\right) + \dfrac{q}{2 l} \mathbf{g},
\end{equation} 
with $\mathbf{l}$ defined as before. 
The translational invariance of $\mathbf{E}$ in \eq{transe} is another reason in favour of our definition of the electric field as $E^{i} = F^{i}_{\phantom{i}k} u^k$.

\section{Weight of the electrostatic energy}

The results obtained above lead to an interesting consequence when we consider the forces
exerted by two charges --- located in a weak gravitational field --- on each other \cite{sumofforces}. To provide
a concrete realization of this situation, consider the following thought experiment.  Two 
charged particles of masses $m_1$ and $m_2$ and 
charges
$q_1$ and $q_2$
are held supported in a weak gravitational field by, for example,
hanging the two particles by strings attached to the ceiling of a room in Earth's gravitational
field, so that the charges are located on the same horizontal plane. If the particles were uncharged
then the sum of the  tensions on the two strings will be equal to the total weight of the particles,
$(m_1+m_2)g$. 
When the particles are charged, they exert electrostatic forces on one another. If we ignore the effect of gravity on the electrostatic 
field produced by the charges, then the force exerted by the charges on one another
is the usual Coulomb force which is directed horizontally along the line 
joining the charges. These Coulomb forces cancel each other and there is no \textit{net} electrostatic
force acting on the charges. 

The situation changes in a curious manner when we take into account the distortion 
of the field lines due to the weak gravitational field. From \eq{transe} we find that
there is a component of the electric field in the direction of $-\mathbf{g}$ produced
by each charge at the location of the other. When we add up the forces exerted by the 
two charges on each other, the forces in the direction of $\mathbf{l}$
cancel out leading to the net extra force given by
\begin{equation}
\mathbf{F}_{12} +\mathbf{F}_{21} = - \frac{q_1q_2}{l} \mathbf{g} = \frac{q_1q_2}{l} \mathbf{g}_{\rm e}
\end{equation} 
In the last expression we have used the fact that the direction of acceleration
in the Rindler frame $\mathbf{g}$ and the direction of Earth's gravitational field $\mathbf{g}_{\rm e}$
are opposite to one another. This result shows that the two strings supporting the charges located
in a weak gravitational field have to support an additional weight $(q_1q_2/lc^2) g$ \textit{which
can be interpreted as the weight of the electrostatic potential energy}. In fact, we 
can turn this argument around to claim that the distortion of the electric field due to 
gravity \textit{must} produce a term of the form $(q/l) \mathbf{g}$ since gravity has to support the electrostatic energy. Obviously, the result 
extends to any number of charged particles all located in the same horizontal plane;
the extra weight that needs to be supported by the string will be equal to the 
effective weight of the total electrostatic energy of the system.

Let us next consider a situation in which two charges are located
at arbitrary positions $\mathbf{r}_1$ and $\mathbf{r}_2$ in the Rindler frame, not
necessarily in the same horizontal plane orthogonal to $\mathbf{g}$.
We can again compute the sum of the two forces exerted by the charges on each other.
From the expression in \eq{transe}, we see that when the locations of the charges are 
interchanged, the Coulomb term flips sign while the term involving $\mathbf{g}$ does not.
Hence, we now get the  total force to be
\begin{equation}\label{forcesum1}
\mathbf{F}_{12} + \mathbf{F}_{21} = \dfrac{q_1 q_2}{r}\left(\mathbf{g}_e + (\mathbf g_e \cdot \mathbf{\hat r}) \mathbf{\hat r}\right) 
\end{equation}
where $\mathbf{r} = \mathbf{r}_2 - \mathbf{r}_1$ and $\mathbf{\hat r}$ is the unit vector in the direction of $\mathbf{r}$.
The origin of the extra term proportional to $(\mathbf g_e \cdot \mathbf{\hat r}) \mathbf{\hat r}$ can be understood by studying the same system
of charges in the inertial frame \cite{pageadams}. To maintain the relative position of two
accelerated charges in the inertial frame, it is necessary to exert 
extra forces in the direction of separation of the charges.
When $\mathbf{\hat r}\cdot \mathbf{g}=0$, these forces are orthogonal to the 
velocity of charged particles and do not do any work. But when   $\mathbf{\hat r}\cdot \mathbf{g} \ne 0$,
these extra forces have to be taken into account in the 
energy and force balance leading to the modified expression obtained in 
\eq{forcesum1}. This result can be verified by explicitly computing 
the total force between two uniformly accelerating charges
moving parallel to each other in an inertial frame; we will do that in the next section.

Note that --- 
in this case, when $\mathbf{g}\cdot\mathbf{\hat r}\neq0$ --- we get a different result if we use the  electric field  $\mathbf{E}_{LL}$
and define (incorrectly, as sometimes done in the literature; see e.g.,\cite{pinto}) the force acting on a charge  to be $\mathbf{F}=q \mathbf{E}_{LL}$. 
From the expression in \eq{ell}, we see that interchanging the position of two charges
will flip the sign of the entire first term leaving $(q/2l)\mathbf{g}$ unchanged.
Therefore, in this case, we will get  precisely the weight of the 
electrostatic energy for the sum of the forces acting on the charges  if 
we define them as $q\mathbf{E}_{LL}$.
This is, however,
incorrect because the force acting on a charged particle is $q\mathbf{E}$ and not 
$q\mathbf{E}_{LL}$. 
The correct result for the total force in this particular context
is indeed given by the expression in \eq{forcesum1}.

 \section{Radiation from a charge and radiation reaction }
 
 \subsection{Radiation from an accelerated charge} 
 
We know that a charge at rest in the Rindler frame will be moving in a uniformly accelerated trajectory in the inertial frame. Therefore, by transforming the electric field given in \eq{origine}, 
reproduced here for convenience:
\begin{equation} \label{origine1}
\mathbf{E}_{Rind} = 
 \dfrac{q \hat{\mathbf{r}}}{r^2}\left(1-\dfrac{(\mathbf g \cdot \mathbf{ r})}{2}\right)- \dfrac{q}{2r} \mathbf{g}
\end{equation} 
to the inertial frame, we should be able to obtain both the Coulomb and radiation field of an accelerated charge.
It is important to carry out this exercise for two conceptual reasons:

(a) The radiation field in the inertial frame is proportional to the acceleration and falls inversely with distance.  We see that such a term arises naturally in the Rindler frame in \eq{origine1}, due to the gravitational field having to support the weight $(qg/rc^2)$ of the electrostatic potential. (Of course, there is no radiation in the Rindler frame in the region under consideration because the magnetic field vanishes identically.) It is conceptually interesting to see how this leads to the radiation field in the inertial frame. It is, however, worth pointing out that, in the Rindler frame, the acceleration dependent term is proportional to $[\mathbf{g} + (\mathbf{g}\cdot \mathbf{\hat r})\mathbf{\hat r}]$, while the radiation term in the inertial frame involves the transverse part of the acceleration, $[\mathbf{g}-(\mathbf{g}\cdot \mathbf{\hat r})\mathbf{\hat r}]$, with a crucial relative minus sign, but evaluated at the retarded time. It is interesting to see how this change arises due to the coordinate transformation.

(b) This exercise actually allows us to find the fields of an \textit{arbitrarily moving} charged particle in the inertial frame, not just that of a uniformly accelerated charge \cite{agtp}. To prove this we can argue as follows: We know from the structure of Maxwell equations, written in the form $\square F_{ik}=4\pi (\partial_{i}J_{k} -\partial_{k}J_{i}) $ that $F_{ik}$ can only depend on the position, velocity and acceleration --- but not on higher derivatives of the trajectory --- at the retarded time of the charge we are interested in, which we will call charge A. We now choose our Lorentz frame such that charge A was at rest at the origin at the retarded time, thereby eliminating the velocity dependence (which, anyway, can be brought in at the end by a Lorentz transformation). We then rotate the coordinates so that the acceleration of charge A at the retarded time is along the $x-$axis.
Next, we introduce another \textit{uniformly accelerated} charged particle (charge B) which has exactly the same acceleration as the charge A we are interested in and is at the origin with zero velocity at the retarded time.
Thus both charge A and charge B have identical position, velocity and acceleration at the retarded time and hence will produce identical fields at $x^i$. We can find the field due to charge B (which is a comparison charge moving on a \textit{uniformly accelerated} trajectory) by the coordinate transformation of the field in \eq{origine1}, and thus determine the field of an \textit{arbitrarily moving} charged particle. As was shown in \cite{agtp}, this idea works for the exact fields in \eq{esol}, but results are unwieldy and it is not clear how the radiation term arises. Hence, we will rework it out to the lowest order, for a charge moving nonrelativistically.  
 
We will transform the expression in \eq{origine1} to the inertial frame and show that the resulting expression agrees with the standard formula for the electric field of a charge which moves with a uniform acceleration $\mathbf{g}$. 
We will work at linear order in $g$ using \eq{approxtr} and \eq{approxtr1}. (We easily extend it to higher orders since we know from the work in \cite{agtp} that the exact expressions can be obtained.) 

The following point, however, needs to be noted. We have defined the electric field in the Rindler frame as $E^i = u_k F^{ik}$ where $u_k$ is the four-velocity of an observer at rest in the Rindler frame, and we are transforming this expression to obtain the electric field in the inertial frame. Rigorously speaking, the latter is given by
$E^i = v_k F^{ik}$ where $v_k$ is the four-velocity of an observer at rest in the inertial frame.  However, it turns out that to the lowest order in $g$ at which we are working, the use of either $u_k$ or $v_k$ leads to the same result. Hence, this difference is irrelevant for the following discussion.

The transformation of the electric field is facilitated by the fact that $\mathbf{E}$ is actually \textit{invariant} to the lowest order. To see this, we use \eq{approxtr} to note that:
\begin{equation}
 E^{X} = F^{X0} =  \left(\dfrac{\partial X }{\partial x}\dfrac{\partial T }{\partial t} -  \dfrac{\partial X }{\partial t}\dfrac{\partial T }{\partial x}\right) F^{x0} 
\simeq  (1 + \mathbf{g}\cdot\mathbf{R}) F^{x0} 
=NF^{x0}=E^{x}
\end{equation}
\begin{equation}
E^{Y} = F^{Y0} = \dfrac{\partial T }{\partial t} F^{y0} 
\simeq  (1 + \mathbf{g}\cdot\mathbf{R}) F^{y0} 
=NF^{y0}=E^{y}
\end{equation} 
where one of the equalities is accurate to lowest order in $\mathbf{g}$. (On the other hand, $\mathbf{E}_{LL}$ is not invariant because of the extra factor of $N$.)
Hence no `transformation' of the field is required and simply substituting the transformed co-ordinates into the original expression for the electric field will give us the result in the inertial frame.
To do this, we note that \eq{approxtr1} gives:
\begin{equation}\label{expforr2}
r^2 = R^2 - T^2(\mathbf{g}\cdot\mathbf{R})
\end{equation}
 On substituting Eqs. (\ref{expforr2}) and   (\ref{approxtr1}) into \eq{origine1}, we find the field in the inertial frame as that in the Rindler frame, expressed in terms of the inertial frame co-ordinates:
 
 \begin{eqnarray}
 \label{geniner}
\mathbf{E}_{iner}&=& \mathbf{E}_{\rm Rind} = \frac{q\mathbf{R}}{R^3} - \frac{q}{2R} \mathbf{g} \left( \frac{T^2}{R^2} + 1\right)  + \frac{q}{2R} (\mathbf{g}\cdot \mathbf{\hat R}) \mathbf{\hat R} \left[ \frac{3T^2 }{R^2} - 1\right]\nonumber\\
&=& \frac{q\mathbf{R}}{R^3}- \frac{q}{2R} \left( \frac{T^2}{R^2} \right)\left( \mathbf{g} - 3(\mathbf{g}\cdot \mathbf{\hat R}) \mathbf{\hat R}\right) - \frac{q}{2R} \left( \mathbf{g} + (\mathbf{g}\cdot \mathbf{\hat R}) \mathbf{\hat R}\right)
\end{eqnarray}
It can be shown that (see Appendix \ref{app:inertialstdE}) that this is identical to the standard result one obtains using the textbook expressions for the electric field and retaining the terms to the lowest order as we have done. While this expression by itself is not very illuminating, there are two special cases which are noteworthy. 

Consider first the field on the $T=0$ surface. Since $\mathbf{r}=\mathbf{R}$ on this hypersurface, we should get exactly the same form of the field as in \eq{origine1}, with $\mathbf{r}$ replaced by $\mathbf{R}$. Putting $T=0$ in \eq{geniner}, we get:
\begin{equation}
\label{eminkows1}
\mathbf{E}_{iner} 
= \dfrac{q {\mathbf{\hat R}}}{R^2} - \dfrac{q }{R}[\mathbf{g} +(\mathbf{g} \cdot \mathbf{\hat R}) \mathbf{\hat R}]
\end{equation} 
which is the same as in \eq{origine1}, with $\mathbf{r}$ replaced by $\mathbf{R}$. 

The fields in \eq{geniner} and \eq{eminkows1} are functions of spacetime coordinates $(T,\mathbf{ R})$ rather than  functions of coordinates at the retarded time --- which is the usual manner in which results are quoted in the literature. To see the connection --- and to illustrate the Coulomb and radiation terms without the complication of the velocity terms --- consider the field along the null surface $T=R$ in the inertial frame. The relevant retarded time for fields on this surface is just $T=0$ and a uniformly accelerating charge with trajectory $(1/2)\mathbf{g}T^2$ will be at rest at the origin at the retarded time. Therefore the retarded position vector is just $\mathbf{R}$ and we should obtain the standard textbook form of the result with the right hand sides expressed at the retarded time.
If we put $T=R$ in \eq{geniner}, we get
\begin{equation}
 \label{geniner1}
\mathbf{E}_{iner}
= \frac{q\mathbf{R}}{R^3}- \frac{q}{2R} \left( \mathbf{g} - 3(\mathbf{g}\cdot \mathbf{\hat R}) \mathbf{\hat R}\right) - \frac{q}{2R} \left( \mathbf{g} + (\mathbf{g}\cdot \mathbf{\hat R}) \mathbf{\hat R}\right)
\end{equation}
which on
simplification, leads to:
\begin{equation}\label{eminkows}
\mathbf{E}_{iner} 
= \dfrac{q {\mathbf{\hat R}}}{R^2} - \dfrac{q }{R}[\mathbf{g} - (\mathbf{g} \cdot \mathbf{\hat R}) \mathbf{\hat R}]
\end{equation} 
which is indeed the standard textbook result involving the Coulomb and radiation fields in Lorentz frame in which the charge was at rest at the origin at the retarded time; as we said before, the velocity dependence can be introduced by a Lorentz transformation. It is worth noting how the Coulomb term and the weight of the electrostatic potential combine in \eq{geniner1} to change $[\mathbf{g}+(\mathbf{g}\cdot \mathbf{\hat R})\mathbf{\hat R}]$ to the correct transverse acceleration $[\mathbf{g}-(\mathbf{g}\cdot \mathbf{\hat R})\mathbf{\hat R}]$.

For completeness, we also obtain the magnetic field in the inertial frame which arises, for example, through:
\begin{equation}
F^{XY}=\dfrac{\partial X}{\partial t}F^{0y}=(gT)F^{0y}=gR\dfrac{qY}{R^3}=\dfrac{qg}{R}\dfrac{Y}{R}
=(\mathbf{\hat R}\times \mathbf{E})^Z
\end{equation}
This is again a standard result.

Finally, the result in \eq{geniner} can also be used to verify a claim we made in the last section regarding the sum of the forces acting on two uniformly accelerated charges. Consider two charges A and B moving along the trajectories 
\begin{equation}
\mathbf{Z}_A = \frac{1}{2} \mathbf{g} t^2; \qquad 
\mathbf{Z}_B = \mathbf{L} +\frac{1}{2} \mathbf{g} t^2
\label{ZAB}
\end{equation}
The field of the first charge everywhere in spacetime is given by \eq{geniner} while the field of the second charge can be obtained from \eq{geniner} by replacing $\mathbf{R}$ by $\mathbf{R}-\mathbf{L}$. At any given time $T$ we can now compute the force exerted by A on B and vice versa. A straightforward calculation shows that:
\begin{equation}
 \mathbf{F}_A (\text{on }B) = \frac{q_A q_B}{L^3} \mathbf{L} - \frac{q_Aq_B}{2L} \left( \mathbf{g} + 
(\mathbf{g}\cdot \mathbf{\hat L}) \mathbf{\hat L}\right)
\end{equation} 
and
\begin{equation}
 \mathbf{F}_B (\text{on }A) = - \frac{q_A q_B}{L^3} \mathbf{L} - \frac{q_Aq_B}{2L} \left( \mathbf{g} + 
(\mathbf{g}\cdot \mathbf{\hat L}) \mathbf{\hat L}\right)
\end{equation}
It is clear that the second expression is obtained from the first by flipping the sign of $\mathbf{L}$ and --- somewhat more surprisingly --- 
the force is time independent. Neither result could have been guessed \textit{a priori} because radiation fields and retarded times are involved. The total force acting on the system is given by
\begin{equation}
\mathbf{F}_A (\text{on }B)+\mathbf{F}_B (\text{on }A)
=- \frac{q_Aq_B}{L} \left( \mathbf{g} + 
(\mathbf{g}\cdot \mathbf{\hat L}) \mathbf{\hat L}\right)
\end{equation} 
which is identical to the result obtained in the Rindler frame due to electrostatics alone. Note that because the electric fields are invariant to the order of approximation we are working, the forces are also invariant under the coordinate transformation.

\subsection{Radiation reaction in the nonrelativistic limit}

Finally, we shall consider an intriguing application of the above analysis: that of determining the \textit{radiation reaction force} on an accelerated charged particle. We know that a charge with \textit{variable} acceleration
will feel a radiation reaction force in the inertial frame proportional to $\dot g$ in the nonrelativistic limit. In the last section we argued (based on \cite{agtp}) that the electromagnetic fields of this charge --- with \textit{variable} acceleration --- can actually be determined from knowing only the fields of a uniformly accelerated charge in the Rindler frame. The question arises as to whether we can also interpret the radiation reaction in the Rindler frame.
It was demonstrated in \cite{agtp} that this is indeed possible but again, since a fully relativistic derivation was given, the actual origin of the radiation reaction force was somewhat obscure. We will now rederive this result in the nonrelativistic limit in a more transparent manner as follows:

We know that a charged particle which has a uniform acceleration $\mathbf{g}$ in the inertial frame can be mapped to a charged particle at rest at the origin of the Rindler frame. The electric field produced by this charge in the Rindler frame is given by \eq{origine} which is accurate to lowest order in $g$. Since the Rindler frame is a static frame of reference, we can, without loss of generality, choose to measure this field at the time $t = 0$.

Let us now suppose that the charged particle is at the origin of the inertial frame (which coincides with the origin of the Rindler frame) at $t=T=0$, but its acceleration $g$  is slowly varying in time with a small but non-zero time derivative, $\dot g$. In other words, the instantaneous acceleration of the charged particle at any time $t$ (near $t = 0$) can be expressed as $g(t) \approx g_0 + \dot g t$ where $g_0$ is a constant and $\dot g$ is small and higher derivatives ($\ddot g$, $\tdot g$ etc.) are ignored. The trajectory of this charged particle can now be expressed in the Rindler frame. The charge is no longer located \textit{always} at the origin, but has a trajectory given by $x_0(t) = \dot g t^3/6$. So, the position of the particle in the Rindler frame now changes with time due to the time derivative of the acceleration $\dot g$.

It is precisely this case that we are interested in for the radiation reaction calculation. We will now to derive the expression for the electric field of a charged particle which moves with slowly varying $g$ as described above, retaining only terms to lowest order in $g$ throughout the analysis: First, we will obtain the expression for the electric field, along the $x-$axis, of a charge that is at rest at the origin of the Rindler frame. Then, we will modify this expression for the case of a charge that is not exactly at rest, but has a small but non-zero $\dot g$. To the lowest order of approximation, this can be  accomplished by replacing $g$ everywhere in the electric field expression, by $g(t) = g_0 + \dot g t$, and at the same time replacing $x$ by $x - \dot g t^3/6$. This latter replacement is necessary because our electric field expression gives the field at point $x$, produced by a charge located at the origin. Since the charge now has the trajectory $x_0(t) = \dot g t^3/6$, the translational invariance of the field requires the replacement of $x$ wherever it appears in the electric field expression, by $x - x_0 = x - \dot g t^3/6$.

We will now carry out the above procedure. Consider the electric field in the Rindler frame of a charged particle which is at rest at the origin of this frame along the $x-$axis of the Rindler frame. 
This electric field is given by setting $r = x$ in the general expression in \eq{origine} leading to:
\begin{equation}
E_x = \dfrac{q}{x^2} - \dfrac{qg}{x}; \qquad E_y = 0
\end{equation}
Replacing $g$ by $g_0 + \dot g t$ and $x$ by $x - \dot g t^3/6$, we get the field due to a charge with a slowly varying acceleration:
\begin{equation}\label{eradreac}
E_x = \dfrac{q}{(x -\dot g t^3/6) ^2} - \dfrac{q(g_0 + \dot g t)}{(x -\dot g t^3/6)}
\end{equation}
This expression is, in general, time-dependent and has to be evaluated at the retarded time corresponding to the field point $x$. Again, to the lowest order  of approximation, the exact nature of the curved path of the null ray does not matter and it can be approximated by a straight line connecting the point $(t,x)$ with approximately the origin (since $\dot g$ is small). Hence, we have $x^2 = t^2$, since the path of light is a null line connecting the above two points. However, since we are measuring the fields at the point $x>0$, say, at the time $t = 0$, the retarded time is negative with $t = -x.$ Effecting this substitution in \eq{eradreac} and retaining terms to lowest order in $g$, we obtain (what will turn out to be) a  miraculous result:

\begin{equation}
E_x = \dfrac{q}{x^2} - \dfrac{qg_0}{x} + \dfrac{2}{3} q \dot g
\end{equation} 

This expression, in the limit of $x\to 0$ is \textit{identical to the expression for the self-force on a charge obtained by Dirac \cite{dirac} with exactly the same coefficients, relative signs and the nature of divergent terms!} The first two terms are well-known  divergences when $x \to 0$, (and are discussed extensively in the literature). Briefly, the first term is discarded as the electrostatic self energy and the second term,  when moved to the left hand side of the equations of motion, leads to a mass renormalization because it is proportional to the acceleration. It is interesting that, even with all our approximations --- working things out to only the lowest order in $g$, and neglecting all higher powers of $g$ throughout the analysis --- we obtain these two terms with their appropriate signs and the correct coefficient factors in front. The real strength of our simple technique, however, is brought out by the production of the last term which is \textit{identical to the standard expression for the radiation reaction field} of a charged particle. (The radiation reaction \textit{force} will be $q$ times this \textit{field}, $(2/3)q^2\dot g$.) Again, the factor and sign in this term are identical to those in the standard expression.
This computation of the radiation reaction  illustrates the  power of our  simple non-relativistic approximation to the electric field. 

\section{Conclusion}
We have discussed several features of electrostatics of point charges in a weak gravitational field. Using a physically motivated  definition for the electric field in curved spacetime, we have shown that one can obtain clear and consistent results and resolve many of the contradictions noted previously in the literature, with respect to the weight of the electrostatic energy, the sum of the forces between charges, etc. We have also proved that the elecrostatic potential of a charge at rest in a weak gravity field, which is usually given as a complicated function of the co-ordinates, can be expressed simply as $A_0 = q/\lambda$ where $\lambda$ is the affine distance along a null geodesic from the charge to the field point. Further, the radiation field of a uniformly accelerated charge in the inertial frame arises as a natural consequence of the transformation of an electro\textit{static} field in the Rindler frame, to the inertial frame. More importantly, we are able to understand the origin of the \textit{radiation reaction}, i.e. the self-force of a charge on itself, by modifying our analysis slightly to include an acceleration which varies slowly in time. 
The extension of these results to arbitrary (and in general, curved) spacetimes is under investigation.

\section*{Acknowledgements}

We thank Donald Lynden-Bell and K. Subramanian for helpful discussions. This work was carried out when we were at the Institute of Astronomy (IOA), Cambridge, and the hospitality provided by IOA is gratefully acknowledged.

\appendix

\begin{center}
\textbf{APPENDICES}
\end{center}

\section{Null geodesics in Rindler coordinates}
\label{app:nullgeo}

We start with the generally covariant form of the Hamilton-Jacobi equation for a null geodesic:
\begin{equation}\label{gencovhje}
g^{ik} \dfrac{\partial S}{\partial x^i} \dfrac{\partial S}{\partial x^k} = 0,
\end{equation} 
where $S$ is the action in a metric of the form
\begin{equation}
ds^2 = g^2 \bar x^2 dt^2 - d\mathbf{r}^2
\end{equation}
Since we are interested in the null geodesics in the $xy$ plane in a static metric, we can separate the variables as:
\begin{equation}
S = -\mathcal{E}t + y k_y + S_1(\bar x)
\end{equation} 
where $\mathcal{E}$ is the energy, and $k_y$ is the $y-$component of the momentum  and $S_1(\bar x)$ stands for the term in the action that depends only on $\bar x$.
For  our metric, \eq{gencovhje} can be reduced to quadrature, leading to
\begin{equation}
S = \int ( \mathcal{E}^2 - k_y^2 g^2 \bar x^2)^{1/2} \frac{d\bar x}{g \bar x} + k_y y  -\mathcal{E}t
\end{equation}
To determine the trajectory in the $xy$ plane, we differentiate $S$ with respect to  
 $k_y$, and equate to a constant $y_0$, getting
\begin{equation}
y- y_0 = k_y \int d\bar x  \dfrac{\ g \bar x}{( \mathcal{E}^2 - k_y^2 g^2 \bar x^2)^{1/2}}.
\end{equation}  
With the substitution $k_y g \bar x = \mathcal{E} \cos \theta$, the above integral can be easily evaluated to find
$
y- y_0  = (\mathcal{E}/k_y g) \sin \theta
$
so that the  equation to the null geodesic is:
\begin{equation}\label{eqcircle}
 \bar x^2 + (y - y_0)^2 = R^2 
\end{equation} 
where $R = \mathcal{E}/k_y g$.
This is the equation to a circle with centre at $(\bar x, y) = (0, y_0)$ and having radius $R = \mathcal{E}/k_y g$.

Let  $\lambda$ be the affine parameter of the null geodesic so that the geodesic trajectory $(t(\lambda),x(\lambda),y(\lambda),0)$ satisfies the equation $k^a\nabla_ak^j=0$ with $k^i=dx^i/d\lambda$.
Since nothing depends on $y$, we have one component of the geodesic equation giving $d^{2}y/d\lambda^2 = 0$ with the solution 
$y = \alpha \lambda + y_0$, where $\alpha$ is a constant and $y_0 \equiv y(0)$. Hence, $y$ itself can be treated as an affine parameter along the trajectory. We will now determine $y$ in terms of $t$.
Along  the null trajectory, we have:
\begin{equation}
 g^2 \bar x^2 dt^2 = d\bar x^2 + dy^2 
\end{equation} 
from which we obtain
\begin{equation}
\label{affine1}
g^2 \bar x^2 \left( \dfrac{dt}{dy}\right)^2 = 1 + \left(\dfrac{d \bar x}{dy}\right)^2 
\end{equation} 
However, from \eq{eqcircle} giving the geodesic trajectory, we know that
\begin{equation}
\left( \dfrac{d\bar x}{dy}\right)^2  = \left(\dfrac{y - y_0}{\bar x}\right)^2
\end{equation} 
Hence, \eq{affine1} becomes:
\begin{equation}
g^2 \bar x^2 \left( \dfrac{dt}{dy}\right)^2 = 1 + \left(\dfrac{y - y_0}{\bar x}\right)^2 = \left(\dfrac{R}{\bar x}\right)^2
\end{equation} 
giving
\begin{equation}
t = \dfrac{R}{g}\int\dfrac{dy}{\bar x^2}.
\end{equation} 
With the substitutions $y = y_0 + R \sin \theta$ and $\bar x = R \cos \theta$, the above integral can be evaluated to give:
\begin{equation}
t =\dfrac{1}{g} \log \left(\dfrac{1 + \tan \ (\theta/2) }{1 - \tan \ (\theta/2)} \right) =  \dfrac{1}{g} \log \tan \left( \frac{\theta}{2} + \dfrac{\pi}{4}\right) 
\end{equation} 
Substituting back in terms of the original variables, we find:
\begin{equation}
2 \tan^{-1} (e^{gt}) - \dfrac{\pi}{2} = \sin^{-1}\left(\dfrac{y - y_0}{R}\right)
\end{equation} 
Rearranging and simplifying, we have:
\begin{equation}
\tanh gt = \left(\dfrac{y - y_0}{R}\right) = \dfrac{\alpha}{R} \lambda
\end{equation} 
Hence, the affine parameter turns out to be proportional to  $\tanh gt$. Its normalization is fixed by noting that when $g\to 0$, we would like the affine parameter to become $t$. This gives $\lambda=g^{-1}\tanh(gt)$.

\section{Maxwell's equations for $E^i$ and $B^i$}
\label{app:maxeb}
In this appendix we will derive the expressions for the generally covariant derivatives of the four-vectors $E^{i}$ and $B^{i}$ which are defined in terms of the electromagnetic field tensor $F^{ij}$ by the relations
$E^{i} = u_{j} F^{ij}$, and $B^{i} = (1/2) \epsilon^{ijkl}u_{j}F_{kl}$, where $u^{i}$ is the standard generally covariant four-velocity. We begin by noting that the field tensor $F^{ij}$ can itself be expressed in terms of these four-vectors by the equation:
\begin{equation}
F^{ab} = u^b E^a - u^a E^b + \epsilon^{abcd} u_c B_d
\end{equation}
and the dual tensor is given by:
\begin{equation}
(*F)_{ij} = u_j B_i - u_i B_j - \epsilon_{ijab} u^a E^b
\end{equation}
Taking the dot product of one of the Maxwell equations with $u^i$ and manipulating terms, we get:
\begin{eqnarray}
4\pi J^a u_a &=& u_a \nabla_b F^{ba} = \nabla_b E^a - (\nabla_b u_a) F^{ba}\nonumber\\
&=& \nabla_a E^a - ( \nabla_b u_a) ( u^a E^b - u^b E^a - \epsilon^{abcd} u_c B_d)\nonumber\\
&=& \nabla_a E^a + a_i E^i +( \epsilon^{dcba} u_c \nabla_b u_a) B_d\nonumber\\
&=& \nabla_a E^a + a_i E^i + \Omega^i B_i
\end{eqnarray}
where $a^i=u^j\nabla_ju^i$ and $\Omega^l=\epsilon^{lijk}u_i\nabla_ju_k$.
We have used $u^a\nabla_bu_a=0$ in one of the steps. Similarly, by taking the dot product of $u^i$ with the second pair of the Maxwell equations, we get:
\begin{eqnarray}
0&=&\frac{1}{2} u_b \nabla_a(\epsilon^{abcd} F_{cd}) = \nabla_b B^b - (\nabla_a u_b)(u^{b} B^{a}-u^aB^b - \epsilon^{abcd} u_c E_d)\nonumber\\
&=& \nabla_i B^i + a_i B^i - \epsilon^{dcab} u_c \nabla_a u_b E_d = \nabla_i B^i + a_i B^i - \Omega^i E_i
\end{eqnarray}
These are the equations quoted in the text.

\section{Expression for retarded time $t_R$}

Consider the field event $\mathcal{F}=(0,\mathbf{r})$ and the source event at the retarded time $\mathcal{S}=(t_R,\mathbf{0})$, connected by a null ray. Setting $s^2=0$ in the expression for the interval given by \eq{genssqr1} will allow us to determine $t_R$. In \eq{genssqr1},
we are now interested in the case with $\bar x_1=g^{-1}, y_1=z_1=0, t_1=t_R, t_2=0, \mathbf{\bar r_2}=\mathbf{\bar r}$ for which we get:
\begin{equation}
\label{ssqr}
s^2(\mathcal{F};\mathcal{S})=\rho^2+ \bar x^2+ g^{-2}-2 g^{-1}\bar x\cosh gt_R
\end{equation} 
The condition $s^2=0$ now determines $t_R$ in terms of other variables and we get:
\begin{equation}
\label{coshtr}
\cosh gt_R=\dfrac{g}{2\bar x}[\rho^2+ \bar x^2+ g^{-2}]\equiv \dfrac{\mu}{\bar x}
\end{equation} 
where the last relation defines the variable $\mu$ which has nice geometrical properties. In particular, it can be expressed as
\begin{equation}
\mu=\dfrac{g}{4}(\ell_+^2+\ell_-^2); \qquad\ell_\pm^2=\rho^2+(\bar x\pm g^{-1})^2
\end{equation}
From \eq{coshtr}, we  find that
\begin{equation}
\label{sinhtr}
\sinh gt_R=\dfrac{\sqrt{\mu^2-\bar x^2}}{\bar x}=\dfrac{g}{2\bar x}\ell_+\ell_-
\end{equation} 
Note also that $\tanh(gt_R/2)=( \ell_-/\ell_+)$ and
\begin{equation}
s^2=\ell_-^2\cosh^2(gt/2)-\ell_+^2\sinh^2(gt/2)
\end{equation} 
More explicitly, we have
\begin{equation}
\cosh gt_R=\dfrac{1+g^2\bar r^2}{2g\bar x};\quad 
\sinh gt_R=\dfrac{1}{2g\bar x}\left[ (1+g^2\bar r^2)^2-4 g^2 \bar x^2 \right]^{1/2}
\end{equation} 
Taking the ratio to obtain $\tanh gt_R$ and switching back to $x=\bar x-g^{-1}$, leads to the expression for the potential given in \eq{a0sol}. Alternatively, using \eq{coshtr} and \eq{sinhtr}, we can express the potential as
\begin{equation}
A_0=\dfrac{qg}{\tanh gt_R}=\dfrac{qg}{2}\left( \dfrac{\ell_+}{\ell_-}+  \dfrac{\ell_-}{\ell_+}   \right)
\end{equation} 
This expression shows that the equipotential surfaces are circles in, say, the $xy$ plane, corresponding to $( \ell_+/\ell_-) =$ constant.

\label{app:rettime}

\section{Potential due to charge at rest at an arbitrary point}
\label{app:trantrick}
Consider the Rindler frame having the metric $ds^2 = (1 + \mathbf{g} \cdot \mathbf{r})^2 dt^2 - d\mathbf{r}^2$. for simplicity, we will work in the $z = 0$ plane and define $\mathbf{r}^2 = x^2 + y^2$.
The four-potential $A_0$ of a charge at rest at the origin, as measured at a point with position vector $\mathbf{r} = (x,y)$, is given by:
\begin{equation}\label{potrind}
 A_0 = \dfrac{q}{r}\dfrac{1 + gx + g^2 r^2/2}{(1 + gx + g^2r^2/4)^{1/2}}
\end{equation} Suppose now that the charge is not located at the origin but at a point $\mathbf{r_0} = (x_0, y_0) $ in the Rindler frame. We make a translation of co-ordinates from $\mathbf{r}$ to $\bar {\mathbf{r}} = \mathbf{r - r_0}$ in the Rindler frame. Then the corresponding metric with a redefined time co-ordinate becomes:
\begin{equation}
ds^2 = (1 + \mathbf{g} \cdot \mathbf{r})^2 dt^2 - d\mathbf{r}^2 = (1 + \mathbf{\bar g} \cdot \mathbf{\bar r})^2 d\bar t^2 - d\mathbf{\bar r}^2
\end{equation} where the new metric is defined by
\begin{equation}
\bar {\mathbf{g}} = \dfrac{\mathbf{g}}{1 + \mathbf{ g} \cdot \mathbf{r_0}}
\end{equation} and the new time co-ordinate is given by: 
\begin{equation}
\label{ttbar}
 d\bar t =  dt (1 + \mathbf{ g} \cdot \mathbf{r_0})
\end{equation} 
The metric in the barred coordinates is identical in form to the original Rindler metric metric and hence, the potential due to a charge located at the origin $\mathbf{\bar r}=0$ in these co-ordinates should also be identical in form to the  potential in \eq{potrind} with the replacements $\mathbf{r}\to\mathbf{\bar r}, \mathbf{g}\to
\mathbf{\bar g}.$ Hence:
\begin{equation}
\bar A_0 = \dfrac{q}{\bar r}\dfrac{1 + \bar g \bar x + \bar g^2 \bar r^2/2}{(1 + \bar g \bar x + \bar g^2 \bar r^2/4)^{1/2}} 
\end{equation} 
If we now substitute for the barred variables in terms of the original variables, we obtain:
\begin{equation}
\bar A_0=  \dfrac{q}{\mathbf{|r - r_0|}}\dfrac{1 + g(x + x_0) + g^2(x^2 + x_0^2 + (y - y_0)^2)/2}{( 1 + gx_0) ( 1 + g(x + x_0) + g^2(\mathbf{|r - r_0|})^4/4)^{1/2}}
\end{equation} 
We now transform to the original frame using \eq{ttbar} and note that the charge which was at the origin in barred coordinates is located at $\mathbf{r}=\mathbf{r}_0$ in the original Rindler frame. This gives the final answer \cite{pinto} to be:
\begin{eqnarray}
 A_0 &=& \dfrac{\partial \bar t}{\partial t} \bar A_0 = (1 + gx_0) \bar A_0 \nonumber \\
&=& \dfrac{q}{\mathbf{|r - r_0|}}\dfrac{1 + g(x + x_0) + g^2(x^2 + x_0^2 + (y - y_0)^2)/2}{( 1 + g(x + x_0) + g^2(\mathbf{|r - r_0|})^4/4)^{1/2}}
\end{eqnarray}

\section{Field of a uniformly accelerated charge in inertial frame}
\label{app:inertialstdE}
We shall obtain the electric field (in the inertial frame) of a charge moving along the trajectory $\mathbf{Z}(T)=(1/2)\mathbf{g}T^2$ to the lowest order in $g$. We start with the standard textbook expression \cite{TPvol1} for the electric field of an arbitrarily moving charge, retaining terms to only up to linear order in acceleration and velocity (since $\mathbf{V}=\mathbf{g}T$ is linear in $g$):
\begin{equation}
\label{minkg}
 \mathbf{E} = \left[ \dfrac{q \mathbf{\hat n}}{R^2{}} \left( 1 + 3\mathbf{v}\cdot \mathbf{\hat n}\right) - \dfrac{q\mathbf{V}}{R^2} + \frac{q}{R} \left( \mathbf{\hat n} \times (\mathbf{\hat n} \times \mathbf{a})\right)\right]_{\rm ret}
\end{equation} 
in which all the terms on the right hand side are evaluated at the retarded time $T_R$ which is determined by the condition $(\mathbf{X}-(1/2)\mathbf{g}T_R^2)^2=(T-T_R)^2$. This reduces to
\begin{equation}
 T^2 + T_R^2  - 2 TT_R = R^2 - T_R^2 ( \mathbf{g \cdot R})
\end{equation} 
Solving this, we get, to the same order of accuracy:
\begin{equation}
 T_R = ( T- R) \left[ 1 + \dfrac{1}{2} (\mathbf{g}\cdot \mathbf{\hat R} )(T-R)\right]
\end{equation} 
We next obtain several other relevant quantities evaluated at the retarded time to be:
\begin{equation}
 R_{\rm ret} = T-T_R = R - \dfrac{1}{2}  (\mathbf{g}\cdot \mathbf{\hat R} )(T-R)^2
\end{equation} 
\begin{equation}
 \mathbf{\hat n}_{\rm ret} = \dfrac{ \mathbf{R}_{\rm ret}}{R_{\rm ret}}
= \mathbf{\hat R} - \dfrac{\mathbf{g}}{2R} ( T-R)^2 + \dfrac{\mathbf{\hat R}}{2R} (\mathbf{g}\cdot \mathbf{\hat R} )(T-R)^2
\end{equation} 
\begin{equation}
 \dfrac{\mathbf{\hat n}_{\rm ret} }{\mathbf{R}^2_{\rm ret}}
= \dfrac{\mathbf{\hat R}}{R^2} - \dfrac{\mathbf{g}}{2R^3}(T-R)^2 + \dfrac{3\mathbf{\hat R}}{2R^3}(\mathbf{g}\cdot \mathbf{\hat R} )(T-R)^2
\end{equation} 
and
\begin{equation}
 3 \mathbf{v}\cdot \mathbf{\hat n}\big|_{\rm ret} = 3 (T- R) ( \mathbf{g}\cdot \mathbf{\hat R})
\end{equation} 
where all the expressions are correct to linear order in $g$.
We can now compute the various terms in \eq{minkg} by direct substitution. The first term, for example, becomes:
\begin{equation}
 \dfrac{\mathbf{\hat n}}{R^2}( 1+ 3 \mathbf{v}\cdot \mathbf{\hat n}) \bigg|_{\rm ret}
= \dfrac{\mathbf{\hat R}}{R^2} - \dfrac{\mathbf{g}}{2R^3}(T-R)^2 + \dfrac{3\mathbf{\hat R}}{2R^3}(\mathbf{g}\cdot \mathbf{\hat R} )(T^2-R^2)
\end{equation} 
while the second, velocity dependent, term is
\begin{equation}
 -\dfrac{\mathbf{v}}{R^2}\bigg|_{\rm ret} = - \dfrac{\mathbf{g} T_R}{R^2} = - \dfrac{\mathbf{g}(T-R)}{R^2}
\end{equation} 
Finally, the radiation term gives:
\begin{equation}
 \dfrac{1}{R}  \left( \mathbf{\hat n}(\mathbf{\hat n}\cdot \mathbf{g}) - \mathbf{g}\right)\bigg|_{\rm ret} = \dfrac{1}{R} \left( \mathbf{\hat R}(\mathbf{\hat R}\cdot \mathbf{g}) - \mathbf{g}\right)
\end{equation} 
Putting all these together, we get the electric field in the inertial frame at an arbitrary location $(T,\mathbf{R})$ due to the charge moving along $\mathbf{Z}(T)=(1/2)\mathbf{g}T^2$, correct to linear order in $g$ to be:

\begin{eqnarray}
 \mathbf{E}_{\rm iner} &=& \dfrac{q\mathbf{R}}{R^3} - \dfrac{q}{2R} \mathbf{g} \left( \dfrac{T^2}{R^2} + 1\right)  + \dfrac{q}{2R} (\mathbf{g}\cdot \mathbf{\hat R}) \mathbf{\hat R} \left[ \dfrac{3T^2 }{R^2} - 1\right]\nonumber\\
&=& \dfrac{q\mathbf{R}}{R^3}- \dfrac{q}{2R} \left( \dfrac{T^2}{R^2} \right)\left( \mathbf{g} - 3(\mathbf{g}\cdot \mathbf{\hat R}) \mathbf{\hat R}\right) - \dfrac{q}{2R} \left( \mathbf{g} + (\mathbf{g}\cdot \mathbf{\hat R}) \mathbf{\hat R}\right)
\end{eqnarray}
This is the expression we used in the text.


\begin{thebibliography}{99}

\bibitem{agtp} Abhinav Gupta and T.Padmanabhan, Phys. Rev. D, \textbf{57}, 7241 (1998); [arXiv:physics/9710036]
\bibitem{rindler} W. Rindler, Am. J. Phys., \textbf{34}, 1174 (1966). 
\bibitem{radiationacc} See, for example, T. C. Bradbury, Ann. Phys. \textbf{19}, 323 (1962);
D. G. Bouleware, Ann. Phys. \textbf{124} (1980), 169; R. Peierls, \textit{Surprises in Theoretical Physics} (Princeton University Press, Princeton, 1979), pp. 169 - 188; T. Fulton and F. Rohrlich, Ann. Phys., \textbf{9}, 499 (1960). These papers have references to previous work.
\bibitem{rindpot} E.T. Whittaker, Proc. Roy. Soc. Lond. A \textbf{116}, 720 (1927);  F. Rohrlich, Ann. Phys., \textbf{13}, 93, (1961); G. N. Plass, Revs. Mod. Phys., \textbf{33}, 37, (1961);
H. Bondi and T. Gold, Proc. Roy. Soc. \textbf{A229}, 416 (1955); 
 M Born, Ann.  Physik., \textbf{30}, 1 (1909);
M. H. L. Pryce, Proc. Roy. Soc. \textbf{A168}, 389 (1938); F. Rohrlich, Nuovo Cimento, \textbf{21}, 811 (1961).
 
\bibitem{gron} E. Eriksen and O. Gron, Ann. Phys. \textbf{313}, 147 (2004).
\bibitem{sumofforces} It appears that this problem was first tackled by Enrico Fermi in 
E. Fermi, Nuovo Cimento \textbf{22}, 176 (1921); reprinted in \textit{Enrico Fermi, Collected papers (Note e memorie)} (Chicago
University Press, Chicago, 1962); English translation in \textit{Fermi and Astrophysics}, edited by
V.G. Gurzadyan and R. Ruffini (World Scientific, Singapore, (2007)).
 Subsequently, there have been several papers the results of which did not always agree with each other. See e.g., 
T. H. Boyer, Am. J. Phys. \textbf{46}, 383 (1978); D. J. Griffiths and R. E. Owen,  Am. J. Phys.\textbf{ 51}, 1120 (1983) and the work cited below in \cite{pinto}. These papers also contain more extensive bibliography.

\bibitem{pinto} Pinto, F., Phys. Rev. D \textbf{73}, 104020 (2006).
\bibitem{comment1}
The Rindler metric in \eq{alt1} is conformal to a metric for which the spatial section is a Poincare half-plane. Since the Poincare half-plane is known to have circles as geodesics, this result is obvious. 

\bibitem{LL} 
Landau L.D., Lifshitz E.M. - Vol. 2. \textit{The Classical Theory of Fields}, Butterworth-Heinemann, (4th edition), 1975, p. 257. 

\bibitem{TPvol1} 
 J.D. Jackson, \textit{Classical Electrodynamics}, 3rd ed. (Wiley, New York, 1999), Chap. 14; Padmanabhan, T., \textit{Theoretical Astrophysics
Volume I: Astrophysical Processes}
 (Cambridge University Press, 2000), Chap. 4.

 \bibitem{pageadams} There is again extensive literature on this topic not all of which reaches the same conclusion. (It appears that the earliest work to note the fact that classical electromagnetism violates Newton's third law was J. J. Thomson, Philos. Mag. (5) \textbf{11}, 229 (1881).) The explicit inertial frame calculations are done in L. Page and N. I. Adams, Jr., Am. J. Phys, \textbf{13}, 141 (1945); 
D. Griffiths and E. Szeto,  Am. J. Phys. \textbf{46}, 244 (1978) but using an \textit{approximate} radiation formula involving a a Taylor series in $(T-T_R)$. These papers have references to earlier work.


\bibitem{dirac} P.A.M. Dirac, Proc. Roy. Soc. \textbf{A165}, 199 (1938); F. Rohrlich, \textit{Classical charged  particles} (Addison - Wesley, Reading, MA, 1965).







\end{thebibliography}
\end{document}